\documentclass{aa}  
\usepackage{graphicx}
\usepackage{txfonts}
\usepackage{changes}
\usepackage{natbib}
\usepackage{pdflscape}
\usepackage{xcolor}	
\usepackage{longtable}
\usepackage{subcaption}
\usepackage{tabularx}

\usepackage{hyperref}
\usepackage{cancel}
\hypersetup{colorlinks=true,linkcolor=blue,citecolor=blue,filecolor=blue,urlcolor=blue}
\newcommand\Thbb{$T\rm_{HBB }$}

\newcommand{\Msun}{\ensuremath{\, {M}_\odot}}

\newcommand{\wCen}{$\omega\,$Cen}

\newcommand{\tdelay}{t$_{\rm delay}$}

\begin{document} 

   \title{Dating N--loud AGNs at  high redshift}
   \subtitle{GS\_3073 as a snapshot of a $\omega$\,Centauri--like evolution of a nuclear star cluster }
   \authorrunning{F. D'Antona et al.}
   \titlerunning{The N-enhanced emission of AGN at high z  and the role of AGB stars }

   \author{F. D'Antona\inst{1},  P. Ventura\inst{1},  A. F. Marino\inst{2}, A. P. Milone\inst{2,3},   E. Vesperini\inst{4},  F. Calura\inst{5,6},   M. Tailo\inst{2,1},  R.  Valiante\inst{1}, 
   V. Caloi\inst{7},  A. D'Ercole\inst{6},  F. Dell'Agli\inst{1} 
          }

   \institute{
              Istituto Nazionale di Astrofisica, Osservatorio Astronomico di Roma, Via Frascati 33, 00077 Monte Porzio Catone, Italy 
             \and Istituto Nazionale di Astrofisica -- Osservatorio Astronomico di Padova, Vicolo dell'Osservatorio 5, Padova, I-35122             
              \and Dipartimento di Fisica e Astronomia ``Galileo Galilei'', Univ. di Padova, Vicolo dell'Osservatorio 3, Padova, I-35122
            \and Department of Astronomy, Indiana University, Swain West, 727 E. 3rd Street, IN 47405 Bloomington (USA) 
            \and Dipartimento di Fisica e Astronomia Augusto Righi, Universit\`a degli Studi di Bologna, Via Gobetti 93/2, I-40129, Bologna, Italy     
           \and INAF -- OAS, Osservatorio di Astrofisica e Scienza dello Spazio di Bologna, via Gobetti 93/3, I-40129 Bologna, Italy
  \and Istituto Nazionale di Astrofisica -- Istituto di Astrofisica e Planetologia Spaziali, Via Fosso del Cavaliere 100, I-00133, Roma, Italy 
             }


 
  \abstract
{In this paper we address two major questions raised by recent  {\it James Webb Space Telescope} observations of the young Universe, namely: 1) what are the seed initial masses, and how rapidly have supermassive black holes (BHs) with masses of 10$^6-10^8$\Msun\ grown in active galactic nuclei (AGN) hosted by very young galaxies? 2) What are the plausible explanations for the  super solar abundances of nitrogen in a fraction of young galaxies at high redshift, both with and without evidence of a massive central black hole? While  the redshift itself, in the still forming high redshift galaxies with AGN, is an upper limit to the ages of the seed black holes, any independent  age determination provides a more stringent constraint on the BH accretion modalities and on the initial seed mass.
We focus mainly on the system GS\_3073. This system shows an exceptionally large $\log$(N/O)=+0.42$^{+0.13}_{-0.10} $\ in the gas close to the AGN. We show here that this abundance is consistent with the composition of gas ejected from massive asymptotic giant branch stars. Moreover,  this system shows chemical properties matching those expected at a specific point of the evolution of the abundances in the `extreme' populations of the former nuclear star cluster $\omega$\,Centauri (\wCen). This analogy, along with the N/O, C/O and Fe/O abundances in GS\_3073, lead to an estimate of an age range of 270--440\,Myr for this object, much smaller than the redshift (z=5.5) age of $\sim$1\,Gyr. We also adopt the same criteria to estimate an age for GN-z11.  These two determinations  constrain the BH mass  {\it vs. }age relation:  accretion on the BH must proceed at intermittent super--Eddington rates in the first phases, and at a much lower rate after the first half gigayear of life of the Universe. The intermittency of accretion is also a fundamental requirement to allow the formation of the ``extreme" (N--rich, O--depleted, He--rich) populations today observed in \wCen\ for a large range of metallicities.
 }

 \maketitle
%

\section{Introduction}
\label{sec1}
Supermassive black holes (SMBH) of $\sim 10^9$\,\Msun\ are present in high redshift (z$\sim 6-7$) quasars \citep[e.g.][and references therein]{fan2023}, raising the problem of how such massive objects may form within less than a billion years from the Big Bang. Initial seed black holes (BHs) are expected to be ``light" (masses up to few 100\Msun) or ``heavy" ($>10^4$\Msun).  Massive seeds may be born as a result of a `direct collapse black hole' (DCBH) channel, the collapse of metal free gas clouds at z$\geq$10,
while light seeds may be remnants of pop.III stars, expected to produce BHs of M$>$250\Msun\ thanks to high Jeans mass and low wind mass loss efficiency \citep[e.g.][]{madaurees2001}, or they are the product of dynamical merging of stellar mass BHs in crowded stellar environments \citep[e.g.][]{portegieszwart2002, giersz2015}. Runaway collisions in extremely compact clusters may also produce very massive or supermassive stars that will later on collapse into seed BHs even $>$10$^3$\Msun \citep{fujii2024, laen2025}
Even if light BH seeds were subject to steady Eddington-limited accretion from early times, they cannot grow up to $\sim 10^9$\Msun\ or more at  z$\geq$6
, so it is necessary to resort either to heavy seeds or to phases of super-Eddington accretion \citep[see][for a review]{volonteri2021}. Disentangling the  possible origins and masses of seed BHs requires information on the intermediate steps of the accretion process, examining BHs at higher redshift  \citep[e.g.][]{ferrara2014, inayoshi2020}. This has been made possible today by the {\it James Webb Space Telescope (JWST)}, whose recent observations, extending to much younger ages, detected an unexpected large number of both broad line 
and narrow line 
active galactic nuclei (AGN) \citep[e.g.][]{harikane2023, maiolino2024, chisholm2024}. The properties of the BHs hosted in these AGN, and especially those at z$>$10 and ages of the Universe $<$460\,Myr, may provide key insights into the BH seeds and the origin of SMBHs. 

A small number of high-z young star forming galaxies have been found to be nitrogen rich,  with super solar N/O ratios \citep{cameron2023, marques2023}, all of them compact and showing high interstellar medium densities and high star formation rates \citep{schaerer2024}. 
The high N abundance is not seen at similar metallicities in the nearby Universe \citep[e.g.][]{izotov2012, izotov2023}, nor is predicted by standard chemical evolution models \citep[e.g.][]{vincenzo2016}. 
In a few of these young galaxies, the presence of an AGN is revealed by the study of the diagnostic lines, so these objects apparently correspond to the N--loud AGN, which are a small percentage of the entire AGN sample at all redshifts. 
Interestingly, \cite{isobe2025} find that about a half of the high--z N--rich galaxies have AGN signatures, while probably only a minority of the young non--AGN galaxies are N--rich. Thus N--loudeness looks like a peculiar feature of systems at an early stage of the growth of their central BH. This apparently implies that the high N signature is a rare event in the life of a young star forming galaxy, but it may be a much more common event in the young AGNs life. 

An alternative interpretation of anomalous high N or C content by \cite{rossi2024} is that they are due to  enrichment from faint Pop III SNe, a result obtained incorporating stochastic chemical enrichment from primordial (population III) stars into a chemical evolution model calibrated on the galactic data.\\
Most of these explanations may be apt to describe the present status of the gas in these galaxies, e.g. in the very young Sunburst galaxy \citep{pascale2023}. The main constraint on such models is that we are observing a very short phase (one--few million of years). 
The necessity of reconciling the oxygen abundance with the high nitrogen may also require more than one burst of star formation \citep{kobayashi2024}. The scarce number of young star forming galaxies with high N/O is consistent with the short-lived phases of N--rich gas. 

An anomalous prototype among high--z N--rich galaxies  is  GN-z11, at redshift z=10.6 measured from JWST/NIRSpec \citep{bunker2023}, and having an intrinsic half-light radius of only 0.016$\pm$0.005$\arcsec$ (64$\pm$20\,pc)  in the JADES NIRCam imaging \citep{tacchella2023}. These features suggested that GN-z11 and the other systems with high N/O host a possible Globular Cluster (GC) in formation \citep{senchyna2024, belokurov2023}. The large [N/O] can be consistent with the abundances measured in the ``second generation" stars of GC stars, inducing several groups  \citep[e.g.][]{charbonnel2023, renzini2023, marques2023, dantona2023} to propose that we may be witnessing the second stage of star formation which apparently occurred in all galactic GCs \citep[see][for recent reviews]{gratton2019, milonemarino2022}.
As a matter of fact, a large N/O is  just one of the many signatures of GC multiple populations; in fact GC stellar spectra show many other more peculiar abundance anti-correlations of light elements, testifying that the gas forming second generation stars  has been processed by proton capture reactions at high temperature (T$>$40\,MK), leaving the signatures of full CNO processing and of the Ne--Na and Mg--Al chains \citep[e.g.][]{gratton2019}.\\
In the case of GN-z11, \cite{dantona2023}  noticed that the population synthesis necessary to model the spectrum required a very high star formation rate and a very young age, but this requirement was dramatically dependent on having neglected the contribution to the spectral energy distribution of a central accreting massive BH \citep{dsilva2023},  revealed by the  analysis of AGN type characteristic features of the spectrum \citep{maio2024gnz11}. If the age and star formation constraint is relaxed, the large N/O may result from the flow towards the center of gravity (including the central BH) of low-velocity winds from massive asymptotic giant branch (AGB), which are at the basis of the `AGB model' for the formation of GC multiple populations \citep[e.g.][]{dercole2008}. \\
If a central very massive GC  is present at the point source of GN-z11, and GN-z11 contains a massive BH, like the nuclear star clusters (NSCs) generally hosting the SMBH of most galaxies in the local Universe do, {\it the NSC could play a key role to explain both the anomalous N/O and  the formation of the seed of the central BH}. 
The hypothesis that an initial NSC, further acquiring mass from merging with other clusters and/or by massive gas accretion from the surrounding medium is present may in fact justify at the same time the formation and growth of a central BH, and account for conspicuous winds of massive AGB stars, which are N--rich for a long lifetime ($\sim$100\,Myr, to be compared with the few Myr of the N--rich winds of young massive or supermassive stars), and so they would better justify the percentage of N--loud young AGNs, compared to the small fraction of N--rich star-forming young galaxies without AGN \citep{isobe2025}. 
\\
Scaling of the AGN properties to the case of GN-z11 provides a mass of $\sim 1.6 \times 10^6$\Msun\ for this object and its luminosity ($L_{AGN}$) at present appears to be larger than Eddington luminosity ($L_{E}$), ($L_{AGN}=2-5 \times L_{E}$), raising the question of the seed initial mass, and its growth \citep{maio2024gnz11}.  The possible ages of the system compatible with the presence of AGB N/O rich ejecta are  $\sim$\,70--130\,Myr,  but the short age, 70\,Myr, requires to accrete at a rate 5 times larger than the Eddington rate, while the latter age is compatible with the lower limit of 2 times the Eddington rate  \citep{dantona2023}, values that anyway must be sustained {\it continuously} for the whole period of time. In the  \cite{dantona2023}   model, these phases of  fast accretion occur on a seed that might have grown to mass of 100-1000 $M_{\odot}$, thanks to stellar mergers in the core of the NSC, when accretion of the stellar AGB winds and of the gas surrounding the object becomes important. \\
Such accretion rate values are huge, even if the BH may sustain supercritical accretion flows, in the framework of the slim-disk solution \citep{abramovicz1995}. In fact, most attempts to follow the evolution history of the accreting BH  \citep[see, e.g.][]{schneider2023, trinca2023,trinca2024} find a {\it highly variable modulation} of the accretion rate.  \cite{madau2014} show examples of evolution histories with three main episodes of accretion at rates three times the Eddington rate, lasting 50\,Myr and followed by quiescent periods of 100\,Myr.  \cite{volonteri2015} estimate the duty cycles of intermittent phases of  super-Eddington growth, coupled with star formation via positive feedback, which may account for the early growth of SMBH and the coevolution with the host system. Both from `slim accretion' models and from numerical simulations, they argue  for accretion episodes lasting 10$^3$-10$^5$\,yr, and `flow regeneration periods' of 10$^4$-10$^5$\,yr.  Intermittent accretion is also critically discussed by \cite{gilli2017}.  
Consequently, either the accretion history of GN-z11 SMBH begins at a massive enough initial seed, or its accretion time span ---and its total age--- must be large enough to accomodate enough intermittent super-Eddington accretion phases.

Before addressing more in detail the AGB solution to the observation of N--rich gas in the vicinity of a SMBH in high redshift objects, which is the main hypothesis scrutinized in this work, we must shortly discuss also the possibility that the accretion modality itself is the source of the N--rich gas. Attention has been recently given to modelling stars either directly formed in the SMBH accretion disk, or captured into the disk from the surrounding star cluster, and to their evolution through mass accretion  and subsequent wind ejection of stellar gas mixed down to the stellar core \citep[e.g.][]{cantiello2021, dittmann2021, fabj2025}. Modelling capture of stars in the accretion disk had been originally devised to explore a scenario in which the variability of AGNs is fuelled by the engulfing of such stars by the BH \citep{syer1991}. Later on, application to formation of binary black holes and their merging \citep[e.g.][]{wang2021} has been discussed. Obviously, the possibility to ``grow" the stars formed in the disk up to very large masses, and try to understand the consequences of such kind of evolution on the global properties of the gas surrounding the BH is fascinating. In principle, the balance between the mass lost in winds from the supermassive stars living in the accretion disk and the mass accreted from the same disk, replenishing  hydrogen into the H--burning stellar core through full convective transport, can prolongate for a long time the core H burning phase of these object, dubbed ``immortal" stars. The winds, made up by gas processed through the CNO cycle in the stellar core, could be the source of the N-rich gas seen in the region of the young galaxies close to the AGN. This hypothetical scenario depends on several schematic hypotheses and adjusted parameters, necessary to deal with: i) the structure of the accretion disk especially in the very outer region where the disk becomes gravitationally unstable, fragments and forms stars \citep{collin2001}, in fact \cite{fabj2025} adopt two very different schemes to build the disk structure; ii)
the stellar accretion modelling \citep{cantiello2021}; iii) the structure itself of the accreting star \citep{chen2024, dittmann2025} and iv) the feedback effects produced by the accretion itself. For this latter problem, we have just seen that the most probable scenarios for the SMBH growth require cycles of accretion rates on much shorter timescales than the lifetime of 100\,Myr of stationary AGN accretion envisioned in the models of immortal stars. Finally, in the present work we will deal mainly with an N--loud young AGN with a value of log(N/O) so large that all massive star models have serious difficulties to explain it, so this provides further motivation for the exploration of different models for this phenomenon. We will briefly return to this point in the discussion, while in the rest of the paper we concentrate on the more standard role of the NSC
stars as a source of the N--rich gas. \\
In this work, we discuss then the hypothesis that we are witnessing a phase of the galaxy life during the evolution of the massive AGB stars of the central NSC. If this is the case, in principle we may get {\it an independent and stronger age constraint  than provided by the redshift alone.} 
We began by asking whether there are similar system where we can apply this constraint to efficiently pin down ages of the growing BH and constrain their mass growth with time, and the analysis yielded an unexpected insight into the problem.\\
We look at the N/O and C/O abundances of a few high redshift systems with N--loud AGN and compare them with those of AGB ejecta. The AGB ejecta  location in the plane log(N/O) versus log(O/H)+12  is briefly discussed  in Sect.\,\ref{sec22}. The remarkably large N/O abundance of the system GS\_3073 \citep{ji2024},  is hardly compatible with any of the other chemical evolution models proposed in the literature, but is fully consistent with the composition of {\it pure AGB ejecta}. 
This bears an analogy with the extreme populations in the massive GC $\omega$\,Centauri (\wCen), so we summarize what we may infer on the evolution of NSC including a BH component from the study of  \wCen\  as a possible prototype of these systems (Sect.\,\ref{wcen}).
On this basis, we attribute an age to GS\_3073  and also revise the age attributed to GN-z11 in \cite{dantona2023} (Sect.\,\ref{sec5}).   We discuss the consequences for the accretion modalities based on the BH mass versus age relation in Sect.\,\ref{tgrow}, and summarize the results in Sect.\,\ref{conclusions}.

 \section{Exploring and modelling the Nitrogen-rich gas in  high-z AGN}
\label{sec2}
Two major questions are raised by the recent JWST observations discussed in Sect.\,\ref{sec1}: 1) What are the initial seed masses of the objects eventually evolving to become the SMBH at the center of galaxies and what is their accretion and merging history leading to large SMBH masses in less than a billion year  \citep[e.g.][]{volonteri2010, volonteri2021}?  2) Why do so many of these systems at early ages \citep[a half, according to][]{isobe2025} show larger Nitrogen abundances close to the central BH?

 \subsection{A sample of N--loud high z compact systems containing an AGN }

\begin{table*}
\small
\caption{Data for four N-loud compact systems at high redshift containing a BH}
\centering 
\begin{tabular}{l  c  l l l  l l c}
\hline
  Name               &       z        & logM$_{BH}$ & log(N/O) &  log(C/O)  & log(O/H)+12 && References \\
  GN-z11          &  10.6     &   6.2$\pm$0.3       &  $>$--0.3        & --0.5        &    8.$^{+0.46}_{-0.46} $ & & \cite{isobe2023, maio2024gnz11} \\ 
  GN-z11          &  10.6     &   6.2$\pm$0.3       &  $>$--0.25        & --0.78        &    7.83$^{+0.2}_{-0.2} $ & & \cite{cameron2023, maio2024gnz11} \\ 
   ~~~~~  (AGN)               &             &         &  $>$--0.09      & $>$--0.78        &    8.58$\div$9.23    & & \cite{isobe2023} \\  \\
  CEERS1019 &   8.7  &   6.95$\pm$0.37  &  $<$0.14 & $<-1.04$ & 7.94$^{+0.46}_{-0.31}$ && \cite{isobe2023,larson2023}\\
   ~~~~~  (AGN)                      &              &         &     $>$-0.01     &  $<$--0.36     &  8.23 $\div$ 8.50  && \cite{isobe2023} \\ \\
  GHZ9              &  10.145    &   7.9$\pm$0.3 & --0.08 $\div$0.12  & --0.96 $\div$ --0.45 & && \cite{napolitano2024} \\
  GS\_3073         &    5.5       &  8.2$\pm$0.4   & 0.42$^{+0.13}_{-0.10} $   & $-0.38^{+0.13}_{-0.11}$ &8.00$^{+0.12}_{-0.09}$ && \cite{ubler2023}, \cite{ji2024}    \\                   
       (UV,  fiducial)         & & &&&&\\
 \multicolumn{2}{l}{ GS\_3073  (UV, low)  }                    &                          & $<-0.21$  &   $> -1.01$      & &&\cite{ji2024}  \\
 \multicolumn{2}{l}{(Optical w/o outflow)}                          &                          &  $-1.1^{+0.18}_{-0.20}$ &         & &&\\
 \multicolumn{2}{l}{ (Optical w/ outflow)  }                     &                          & $-0.58^{+0.18}_{-0.20}$  &        & &&   \\
\\
\hline
\end{tabular}
\label{table:1}
\end{table*}

\begin{figure*}
\begin{minipage}{0.99\textwidth}
\resizebox{.5\hsize}{!}{\includegraphics{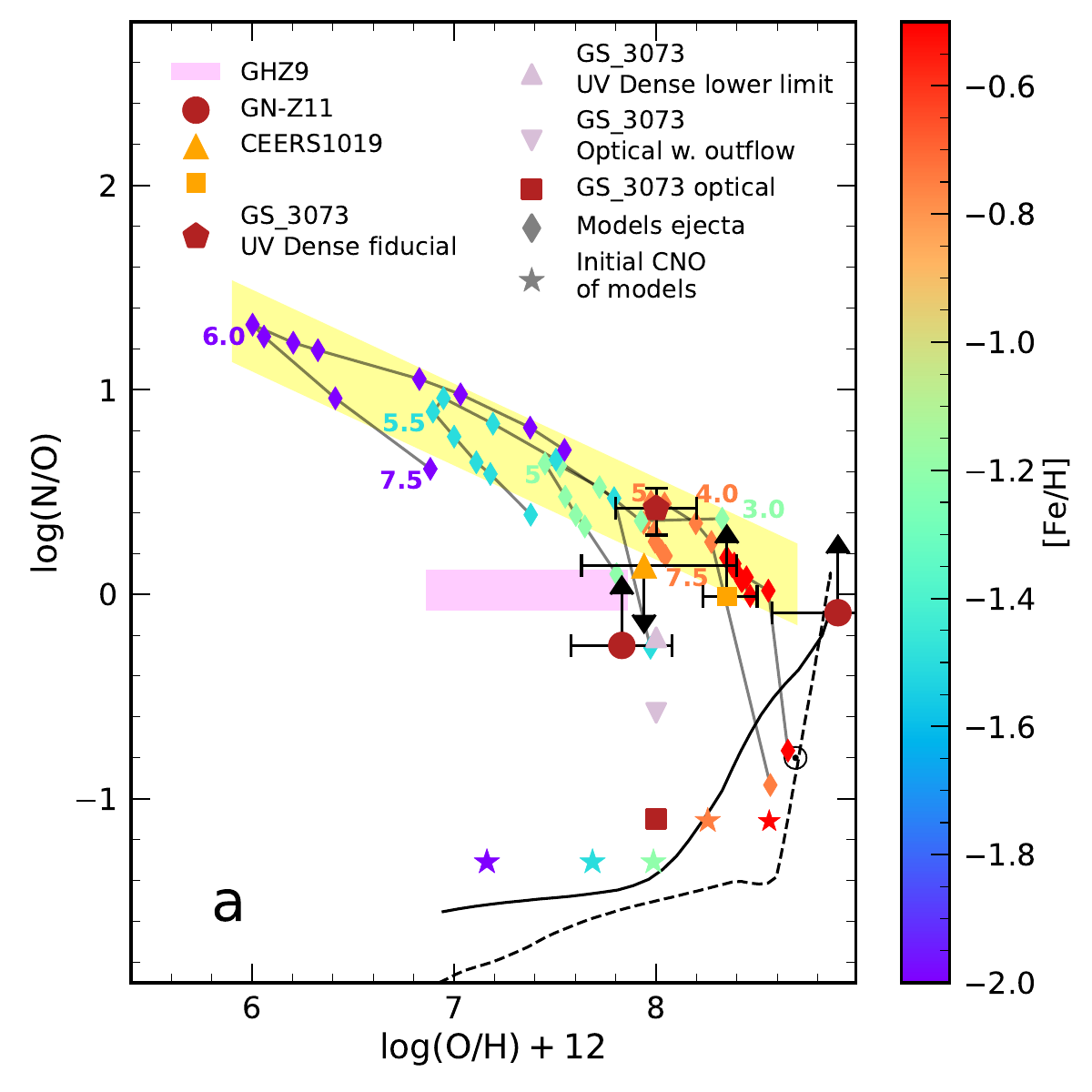}}	  
\resizebox{.5\hsize}{!}{ \includegraphics{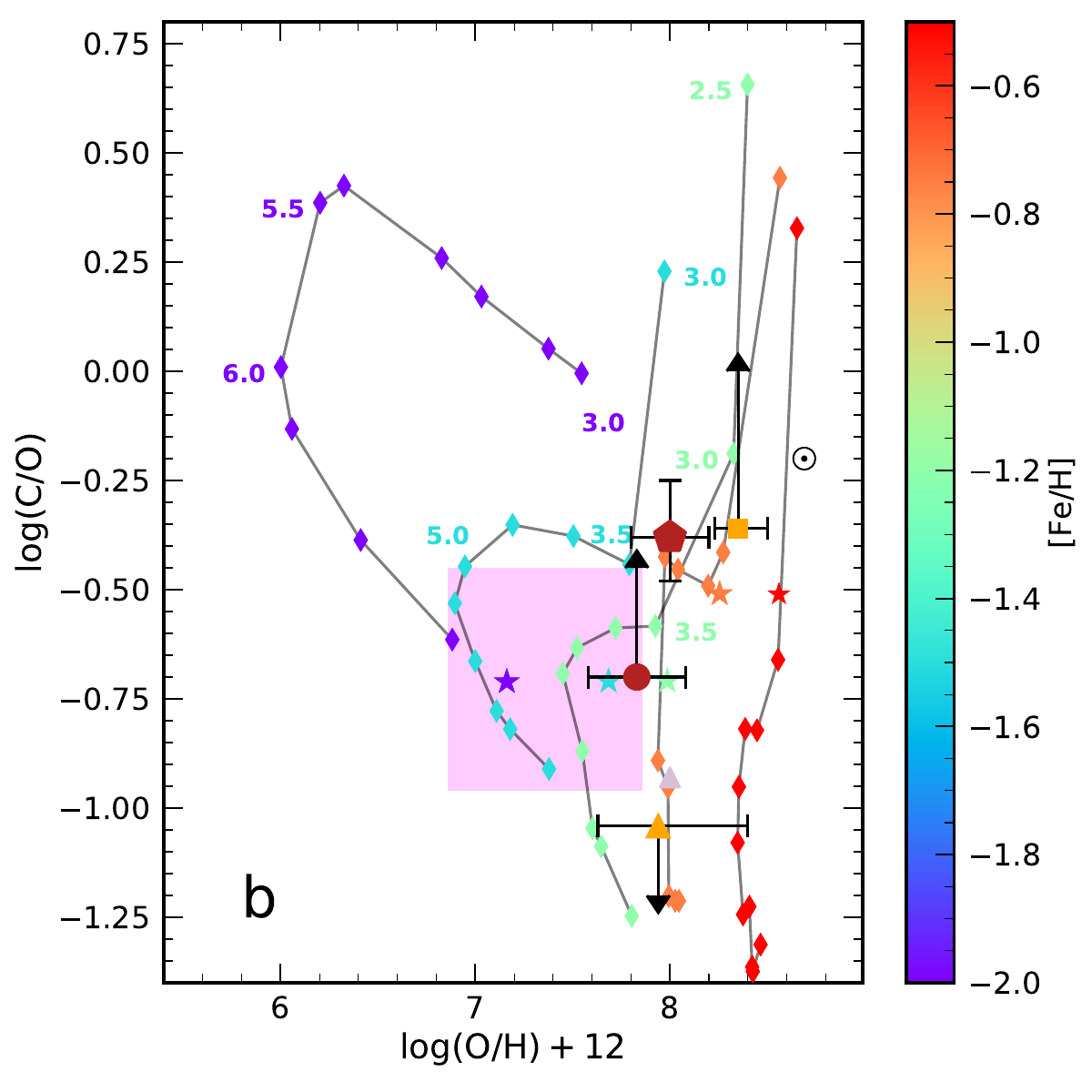}}	 
\end{minipage}
\caption{ Abundances in GN-z11, GHz9, CEERS\,1019 and GS\_3073 compared with models. Panel a: diagram log(N/O) versus log(O/H)+12.  For GN-z11 and CEERS\,1019 we show the standard locations in the literature and the locations resulting from \cite{isobe2023} models including the AGN radiation  (see Table\,\ref{table:1}); for GS\_3073 we show the fiducial (red pentagon) and lower limit (grey triangle) derived from the UV-dense lines, monitoring the region closest to the AGN, while  the log(N/O) from optical lines are the red  square (w/o outflow) and the reverse grey triangle (w/ outflow).
The scale of colour on the right defines the metallicity of the different sets of mass averaged ejecta composition (triangles), for initial   AGB masses from 7.5 to 3\Msun --or 2.5\Msun\ in one case, from the models by \citep{ventura2013}. The metal mass fractions of the tracks Z, and the corresponding [Fe/H] are Z=0.0003 ([Fe/H]=--2), 0.001 (--1.5), 0.002 (--1.2), 0.004 (--0.75) and 0.008 (--0.45). A few masses are also labelled as reference. The yellow band is the reference location of ejecta dominated by hot bottom burning as a function of  log(O/H). Stars represent the  initial O/H and N/O  abundances of the models. The solar symbol is placed at the values corresponding to solar abundances.
At the bottom: upper (full line)  and lower (dashed line) envelope  of the chemical evolution models by \citet{vincenzo2016}.  Panel b: diagram  log(C/O) versus log(O/H)+12. Symbols and lines as in panel a.
}  
\label{fig:1}       
\end{figure*}

In Table\,1 we list four systems sharing the property of (i) high redshift; (ii) high N/O ratio in the spectra; (iii) very probably hosting a SMBH. The systems are listed in order of increasing BH mass.\\
The presence of a SMBH in these systems is very tricky indeed, as the way the abundances are derived depends on whether the emission lines are excited by the stellar radiation or by the harder emission of the AGN \citep{isobe2023}, thus for GN-z11 and CEERS\,1019\footnote{ We include CEERS\,1019 in the list of AGN-hosting galaxies, based on the results by \cite{larson2023} who find that the H$_\beta$ line emission contains a broad component, fitted with a very high rotation velocity, sign of the presence of an AGN Broad Line Region. According to \cite{marques2024} this is a very young star forming galaxy hiding no SMBH.} 
 we show the results for both hypotheses, and have to keep this caveat in mind. \\
The location of GHZ9 \citep{napolitano2024} appears similar to the `standard' location of GN-z11 \citep{cameron2023}, but (O/H) is not determined, so it is plotted with the location corresponding to a plausible interval from 0.01(Fe/H)$_\sun$ to 0.1(Fe/H)$_\sun$.\\
The N/O of two objects, GS\_3073 and CEERS\,1019 (in the AGN spectrum hypothesis), are at values much higher than GN-z11, log(N/O)$\sim$0. and 0.4.  \\
For the most interesting case, GS\_3073, we plot in Fig.\,\ref{fig:1}a both the ``UV dense" fiducial determination (red pentagon), corresponding to the spectrum from the region closest to the BH, its lower limit (grey triangle) and the optical determination (red square) from \cite{ji2024}. The latter value is not different from the average abundances in galaxies at the same level of metallicity, so probably the anomalous high nitrogen is confined to a limited region in the proximity of the AGN. Unfortunately, the dimensions of this region are not known, precluding more quantitative analyses, but not diminishing the interest of the measured abundances. \\
Looking at the location of the four objects in the log(C/O) versus log(O/H) plane (Fig\,\ref{fig:1}b) we point out another possibly interesting feature: the (C/O) of GS\_3073, also considering its lower limit, is quite `normal' with respect to the abundances of the objects plotted in this plane, but the fiducial point is somewhat higher than  the abundance of GN-z11, and it is also large with respect to the bulk of the ejecta abundances of the massive AGBs. \\

\subsection{The limitation to the abundances of AGB ejecta in the plane log(N/O) vs. log(O/H)}
\label{sec22}
In Fig.\ref{fig:1}a we plot the average abundances in the ejecta of AGB stars of different metallicities, from the work by  \citep{ventura2013}, going from Z=0.0003 to Z=0.008\footnote{It is important here to clarify that the evolution of massive AGBs was drastically revised in the latest 20 years starting with the studies by \cite{ventura2001} and the following works, notably by \cite{ventura2005a, ventura2005b, ventura2013}, and we are specifically dealing with this kind of results. When authors are led to reject the role of AGBs in the evolution of these systems, based on previous modellization, their result must be analized with care when they adopted very different yields \citep[see, e.g. the results of ][based on the evolution model by \citeauthor{hamannferland1993} 1993]{isobe2023}. }.
The abundances in the starting main sequence models are represented by the asterisks.  
These average abundances are the result of wind mass loss which progressively (on timescales of about a few$\times10^5$\,yr, depending on the evolving mass) expels from the stellar surface the whole convective envelope until the remnant carbon oxygen core is left and cools off as white dwarf. The structure of massive AGBs are such that their convective envelope reaches the edges of the H--burning shell, and the whole envelope gas is processed at the high temperatures, \Thbb,  of this ``hot bottom" (hot bottom burning or HBB). The \Thbb\ reach progressively values from 40 to 100\,MK, provoking p--capture processing on all the light elements involved. In particular, the C and N abundances reach quasi equilibrium abundances, while  the oxygen conversion to nitrogen goes on for the whole evolution, and the N/O reached is limited by the finite lifetime of the AGB phase  due to mass loss. \\
With respect to the models plotted in Fig.\,1 of \cite{dantona2023}, the initial masses are extended, to 3 or 2.5\Msun, showing the change in the ejecta composition: the upper mass models (7.5--6.5\Msun) are dominated by mass loss, so the oxygen abundance is not significantly reduced, in spite of high \Thbb; the full CNO processing becomes the dominant effect for the 6--5\Msun; for smaller masses, \Thbb\  decreases and oxygen becomes less depleted  again ($\sim$4\Msun); finally the third dredge up \citep{ir1983} becomes dominant, the  NO cycling is no longer efficient,  and the models transition to C-star (about 3\Msun). \\
In the plane log(N/O) versus log(O/H), models of different metallicities are located along a diagonal band tracing the increasing efficiency of CNO cycling as \Thbb\ increases for lower and lower metallicities. The efficiency of HBB is directly dependent on the opacity of the envelope, and thus mainly on the iron content of the mixture. The diagonal band location for our models does not depend on the  $\alpha$\ elements enhancement with respect to the solar ratio\footnote{In the models displayed here it is  $\alpha/$Fe=0.4 for Z$\le$0.002, and $\alpha/$Fe=0.2 for Z=0.004 and 0.008.}, as the $\alpha$\ enhancement  affects the oxygen abundance, and thus it affects in the same way both the abscissa and ordinate of the plot,  and the ejecta shift along the band shaped by the models of different (Fe/H). {\it The diagonal band is then the limit to the abundances we can expect to see in AGB ejecta}. 
Abundances smaller than those in the yellow strip ---such as in the other N/O rich systems we are examining--- can be obtained by a mixture of ejecta with the infall gas abundances. In standard GCs this mixing is with gas having the primordial composition of the cloud forming the first generation, so the mixing line would go from the diamond of the ejecta of the evolving mass to the star of the corresponding  initial composition (in Fig.\,\ref{fig:2}, the star of the same colour). In these complex high-z systems, the metallicity of the infalling matter may be a bit different, e.g. smaller. In any case, the diluting line will go from the strip points down to the stars points, and can not, e.g. go to other points along the strip. An example of the dilution necessary to fit the location of GN-z11 is shown in Fig.\,1 of \cite{dantona2023}. 
\\
The vertical location (the values of N/O for each O/H) is very dependent on the efficiency of  convection in the AGB envelope
and on the mass loss description  \citep{ventura2005a, ventura2005b}.
Our own models are those which best conform to the abundances observed in GCs, and are at the basis of the `AGB model' for the formation of multiple populations \citep{dercole2008, dantona2016, calura2019}, so they are the most appropriate also in this context. 
For example, the models by \cite{karakas2010} and \cite{fishlock2014} adopt lower efficiency of mass loss  with respect to ours, so the AGB evolution proceeds through a much larger number of helium thermal pulses, during which the outer H--rich convective envelope deepens into the helium intershell and ``dredges up" the products of 3$\alpha$ burning, mainly Carbon. Thus the total CNO in the envelope increases dramatically, at variance with the abundances in GCs, where the total CNO varies at most by a factor two between the first and second generation stars.  \\
 For a quick reference, the strip location highlighted in yellow in  Fig.\,\ref{fig:1}a  follows the relation:
\begin{equation}
 \log ({\rm N/O})=(4.05 \pm 0.2)-0.46 \times (\log ({\rm O/H})+12)
\end{equation}
\subsection{Modelling the high Nitrogen in compact objects at high z}
\label{modelsN/Olarge}
The high N/O ratios of the objects listed in Table\,\ref{table:1} and plotted in Fig.\,\ref{fig:1} require some care  to be put into context. 
The first obvious point is that these N/O values are anomalous, they can not be achieved easily in the context of normal chemical evolution models, calibrated on ``standard" signatures of how the elemental abundances evolve in time, taking into account all kinds of stellar polluters (supernovae, winds, mass loss from binaries) together with possible infall  sources (as in the models by \cite{vincenzo2016}  plotted in the figure). So the high N/O of massive stars belong to some anomalous phase, probably of short duration and not influencing the global chemical evolution.\\
One interesting aspect of this problem was shown by Kobayashi and Ferrara (2024): assuming one burst of star formation, high N is obtained from the Wolf Rayet yields, but at the same time oxygen increases too much. So a plausible model invokes a first burst of star formation, followed by a decrease of oxygen thanks to infall of primordial gas, and finally another burst of star formation, some 100--200\;Myr after the first one, to produce again high N,  for a very short time, through the Wolf Rayet channel.  \\
\cite{rizzuti2024} avoid the overproduction of oxygen in the the bursting star formation events, assuming that the fast ejecta of supernovae are lost by the system, while the slower stellar wind products (secondary --and sometimes primary-- nitrogen) remain in the system and pollute the interstellar gas. These models well explain GN-z11.\\
Other models explore the possible role of Very Massive Stars (VMS) \citep{vink2023}. In these stars the winds will be slow in the first phases of evolution and should be able to produce the high N in GN-z11 and in other young star forming galaxies at high redshift.\\
A final attempt is made by \cite{nandal2024}, who resorted to population III stars well above 1000\Msun\ to reproduce the N/O in GN-z11. \\
When we limit ourselves to objects hiding an AGN, the high star formation rate and very young age of the system are no longer mandatory, and also the AGB role can be considered, as discussed in \cite{dantona2023}.

\subsection{The very high N/O in GS\_3073}
An interesting problem arises when we consider the extreme log(N/O)=0.42 of the high density  region of GS\_3073 close to the AGN.  The `formal' full equilibrium values for CNO burning are (log(N/O)$\sim$1, log(C/O)$\sim$--0.3 \citep{isobe2023, maeder2015}, but the average N/O abundance of massive stars  are generally much lower than this \citep[see, e.g., the sketch figure 4 in ][]{isobe2023} --and we have seen above in Sect.\,\ref{modelsN/Olarge} that they meet with difficulty, and for a short time, the much lower GN-z11 abundance ratios. Even the chemical evolutions by \cite{rizzuti2024}, successful in explaining the high N/O in the other objects, had to resort to two peculiar models to reach the N/O in GS\_3073: one with extremely short infall timescale and extreme rotation rates, evolved for 50\,Myr, and another one with a much longer evolution time of 200\,Myr. Note that this latter timescale implies indeed a contribution from AGB winds to the chemical composition of the gas, and in many ways is similar to the hypothesis we develop in the present work. The model with extreme infall corresponds to an exceptionally high star-formation rate, so an accurate measurement of this quantity would be crucial for distinguishing between these two scenarios.  \\
A recent work \citep{nandal2025} shows that the abundances in GS\_3073 can result from the evolution of  1000--10000\Msun\ primordial (population III) stars, under some specific assumptions on the mass lost by these stars, and the precise evolutionary stage at which it is lost. Anyway,  it is unclear how supermassive objects {\it with no metals} could form at such a late stage of evolution, as we are looking at a system containing a SMBH  of $1.6 \times10^8$\Msun\ at z=5.5. For example, hydrodynamic simulations show that, within an individual galaxy, the metal production and stellar feedback from Pop II stars overtake Pop III stars in 20-–200 Myr, depending on galaxy mass \citep{muratov2013}.\\
In the AGBs subject to HBB the evolution of abundances does not reach CNO equilibrium, but we see that the log(N/O) of GS\_3073 is well matched by the AGB ejecta. Values of N/O such as in GN-z11 can be fit by diluting the abundances of the AGB ejecta with infalling gas with standard composition \citep{dantona2023},  a  procedure that is also used to explain the abundances in GC stars of second generation.
In the case of GS\_3073  we seem to be looking at {\it undiluted} ejecta close to the AGN. Is the presence of undiluted AGB ejecta possible and reasonable?  It is possible that the region in the close vicinity of the BH becomes periodically free of all the gas which has been either rapidly accreted on the central BH (as in the intermittent accretion modelling), or has gone through a burst of star formation, so immediately after  each episode of rapid gas depletion the main new contribution to the gas we see comes from the winds from the closest AGB stars, hosted in the putative NSC and evolving at the time we are observing the system.
The accretion rate in GS\_3073 is largely sub--Eddington \citep{ubler2023}, so we might indeed be looking at one of these phases, in which the region closest to the BH is being replenished by the AGB gas.\\
From a more general point of view, we know that in standard GCs the second generation contains, in some cases, a fraction of stars born from pure ejecta: these have extreme p-processing signatures and very large helium content, in mass fraction  Y$\sim 0.35-0.40$. \\
In particular we know that  \wCen\ (likely the remnant NSC of a disrupted dwarf galaxy) probably hosts a number of `extreme' populations with a wide range of metallicities \citep[e.g.][]{johnson2009, marino2011wcen, bellini2017, clontz2025}, supporting the  idea of recurring star formation phases during which stars form out of  only pure AGB ejecta  in the central regions of the system \citep[e.g.][]{dantona2011wcen}.

\subsection{ Formation of multiple populations in standard GCs and in NSCs based on the AGB model}
\label{models}
We are adding together two very important potential features of some compact high-z objects including an AGN component: 1) an NSC in the centre of a more extended galaxy like GN-z11 is the right place where to form and grow the BHs seeds of an  SMBH; 2) the stars evolving in the NSC, after only 40\,Myr, can become sources of high-N gas.\\ 
Let us consider the fate of the winds of the evolving AGBs:  a part of them probably takes part in new events of star formation, like it occurs in standard GCs for the formation of the ``second generation" stars \citep[see, e.g.][]{dercole2008}. Stars in standard GCs, however, are characterized by a small Fe and age spreads.  NSC evolution is qualitatively different, as they host many populations differing in metallicity and age \citep[see the review by][]{neumayer2020AAR}.  The extended star formation history in NSCs may be reconstructed  by fitting their integrated spectrum \citep{walcher2005, kacharov2018}, and by studying the single stars. 
The NSC formation occurs thus on longer lifetimes than the standard GC formation, and has the advantage of allowing for massive accretion of gas from the outside, which help to build up many stellar populations. Unlike in standard GCs, matter ejected by exploding supernovae in NSC is not expelled, thanks to the high escape velocity, and, together with infalling matter, it allows the formation of several subsequent generations with increasing metallicity, while the gas also accretes on the core SMBH. Complications to consider include the gas ejection in jets from the SMBH, and the possible merging with other systems, including additional stars populations and gas.\\
If the formation of the populations in an NSC may be very extended in time, we probably have to revise the assumption in \cite{dantona2023} for the evolution of GN-z11.  We assumed that the most massive stars of the `first generation' of the NSC progenitor  ended their life as the BHs seeds of the SMBH, and that the  AGB winds {\it of the stars of this same generation}  were responsible for the specific composition seen in the spectrum, so the age of these AGBs would also be the total age of the system, fixing it to $\sim$70--130\,Myr.\\
This is not the only possible conclusion. We can consider another case: 1) the seed BHs were born in a first very metal poor population; 2) we are instead looking at the winds from the stars born in more recent events of star formation, in the gas metal enriched by the local supernovae. Under this scenario, we have to estimate the timescale on which the recent parent stellar generation was born.\\
To gain further insight into the formation and chemical history of high-z objects like GN-z11 and GS\_3073, we can compare these systems with the very well known proxy \wCen, the cluster considered a NSC which has lost the surrounding  galaxy, probably the dwarf galaxy Gaia--Sausage/Enceladus \citep[e.g.][]{lee1999, pfeffer2021, limberg2022}.  
Under the assumption that indeed \wCen\ is a stripped NSC, \cite{limberg2022} estimate the stellar mass of its progenitor to be $\sim 1.3\times 10^9$\Msun. The system also harbours a central BH with an estimated mass of $\sim$8000\Msun\ \citep{haberle2024}: in summary, it resembles the initial phases of the high  z systems we are dealing with, in a framework such that the BH growth has been stopped at a very early stage. Also its stellar populations have stopped forming at early times, allowing a better comparison with the early evolution of the high--z objects. 

\section{ \wCen\ as a proxy example}
\label{wcen}
\begin{figure}
\begin{minipage}{0.99\textwidth}
\resizebox{0.5\hsize}{!}{\includegraphics{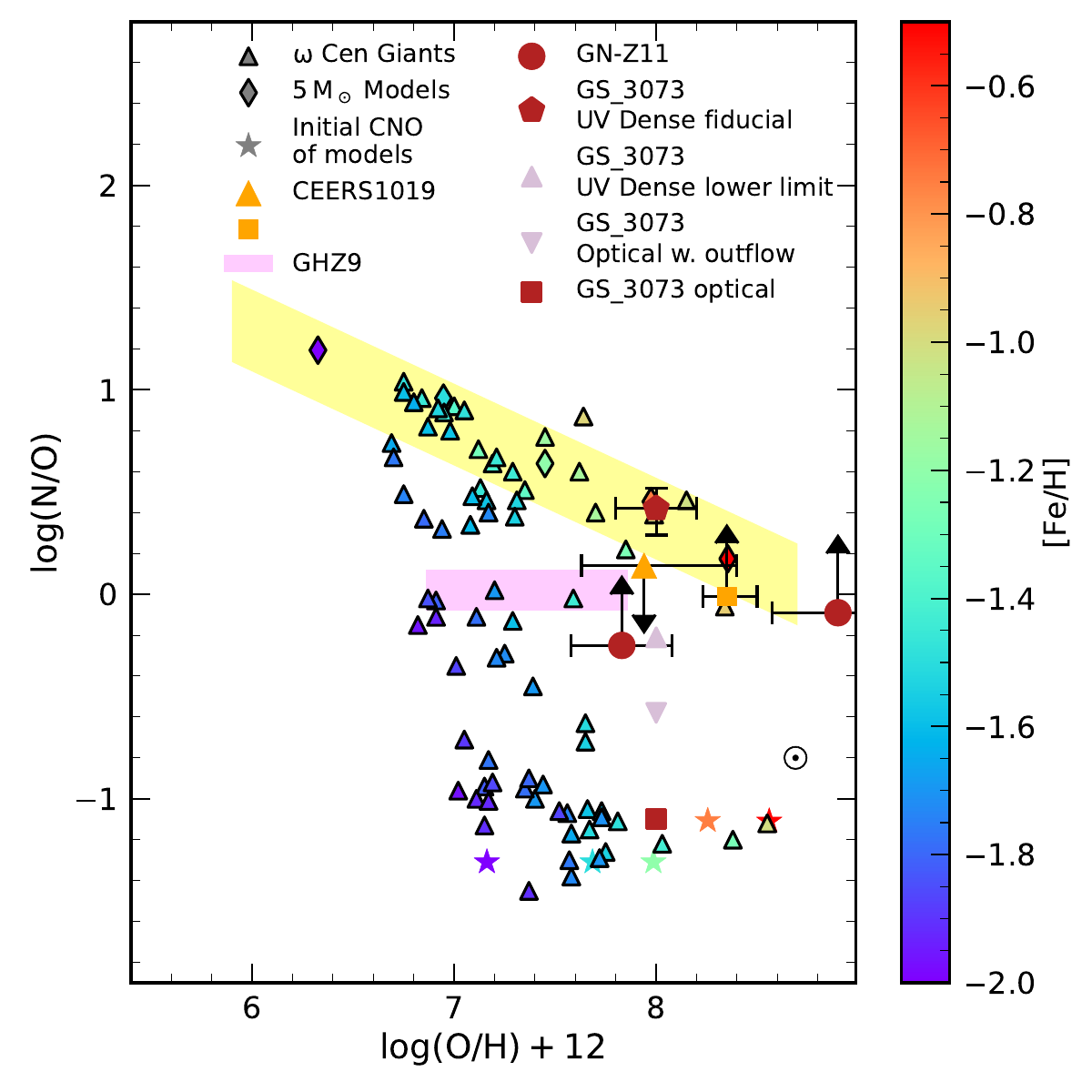}}	  
\end{minipage}
\caption{ In the plane log(N/O) versus log(O/H)+12 we show as triangles the abundances in a sample of \wCen\ giants by \cite{marino2012cno}, for  [Fe/H] according to the scale at the right of the figure. Abundances in GN-z11, GHz9, CEERS\,1019 and GH3073 are displayed as in Fig.\ref{fig:1}.  The yellow diagonal strip represents the locus of pure AGB ejecta of intermediate masses, for a large range of metallicities, defined by the models shown in Fig.\,\ref{fig:1}. The stars are the initial abundances of the evolved models, and the diamonds along the strip are the ejecta of 5\Msun\ of the different sets of tracks. }  
\label{fig:2}       
\end{figure}

A directly appealing reason to consider \wCen\ an example of the early formation history of an NSC
is highlighted in Figure \ref{fig:2},  showing where the giants in \wCen\ are located in the log(N/O)  versus log(O/H)+12 plane. Data for the giants are from \cite{marino2012cno}. Two features immediately emerge: 1) the yellow strip of AGB ejecta composition is a natural limit to the observed abundances; 2) a  fraction of the sample stars, at all metallicities apart from the lowest abundance bin, are actually {\it on the yellow strip}, implying, in the context of the AGB model, that they were born from pure gas ejected by AGBs of the NSC (with a caveat which we will discuss later). It looks like the red pentagon representing the abundances in the dense region of GS\_3073, similarly located on the yellow strip, is a sort of snapshot of the core of \wCen\ at some precise point of its chemical evolution history.  The analogy may serve as a guide  to understand the status  of  GS\_3073.\\
The data shown in Fig. \ref{fig:2} are only one among the evidences that extreme second generation stars form at all metallicities in \wCen, but it is necessary to summarize other important hints, based on other abundance patterns and on the interpretation of photometric data.\\ 
\subsection{Spectroscopic data}
While the metallicity of \wCen\ stars spans much more than a dex, the high dispersion spectroscopy data by \cite{johnson2009} and \cite{marino2011wcen} have shown that the abundances of p--capture elements {\it in each of the metallicity groups} are anticorrelated, with patterns similar to those found in simpler GCs having no spread in metallicity \citep{carretta2009c}. This suggests that \wCen\ passed through a complex evolution, with similarity and differences to the stages forming multiple populations in GCs. The main results were clear: 
\begin{itemize}
\item  the stars in the lowest metallicity bin ([Fe/H]$<$--1.9) show a `mild'  O--Na anticorrelation; 
\item in the intermediate metallicity bins, --1.9$<$ [Fe/H]$<$--1.3, the O--Na anticorrelations (and other anticorrelations) are very extended, implying that a fraction of stars is ``extreme" in abundances, and another large fraction shows abundances intermediate between the standard and extreme abundances; 
\item in the high-metallicity bins  ([Fe/H]$>$--1.3) a large majority of stars has ``extreme" abundances and and are characterized by a O--Na direct correlation\footnote{ \cite{dantona2011wcen} interpreted  these data on the basis of \cite{dercole2008} model, showing that the direct correlation O--Na in the high metallicity bins was expected if in the last phases of star formation the whole populations were made by undiluted  AGB winds of different evolving mass.}.
\item the stars with low Na and high O (identified as  `first generation' stars, as they have abundances similar to the halo stars) are scarce or missing  in the high metallicity bins.
\end{itemize} 
Complementary to these data, Fig.\,\ref{fig:2} shows that the  log(N/O) ratio versus oxygen hosts stars with ``standard" N/O, or with ``intermediate" N/O abundances  (but not the higher metallicity stars), and the stars with the lowest metallicity do not reach extreme N/O  (violet triangles do not populate the yellow strip). The stars with highest metallicities are indeed almost all ``extreme" in N/O, they lie on the ``yellow strip" of pure AGB ejecta.

\subsection{The chromosome map}
\cite{marino2019} displayed the abundances from high dispersion spectra in the `chromosome map' \citep{milone2017chromo} of \wCen\ --where the ordinate pseudo-color represents mainly Nitrogen-- showing that the high dispersion results hold also for a much more abundant (photometric) sample: for most of the metallicities (apart from the lowest ones) a fraction of stars has the highest p-capture abundances,  and the first generation stars are scarce or absent at the high metallicity. 
Similar results have been recently found by \cite{clontz2025}, who matched the chromosome map data with the metallicities determined in low dispersion MUSE spectra for thousands of stars.

\subsection{The photometric evidence}
In \wCen\  standard photometry in the optical bands shows the presence of a ``blue main sequence" \citep{bedin2004},  matched by isochrones with very high helium content, e.g. Y$\simeq 0.39\pm$0.02 \citep{king2012}, or  Y$\simeq$0.37 \citep{tailo2016}. Subsequently, colour magnitude diagrams in the UV HST bands, complemented with the chromosome maps of the stellar groups photometrically identified, allowed \cite{bellini2017} to reveal  more than 15 different populations. Among them, {\it there are in total 8 populations with high helium}: three constituting the blue MS, three in what they call MSd  (hidden beyond the red MS in standard CM diagrams) and  two in the MSa, the very red and metal rich part of the diagram. These result for the main sequence match the finding,  quoted above, of a high helium abundance ($\delta$Y$\sim$0.15 with respect to standard) in all the populations\footnote{Again, the stars with the lowest metallicities have only modest helium enhancements.} of the
upper portion of the giants' chromosome map from the oMEGACat survey  \citep{clontz2025}. \\
\subsection{Modelling the star formation in \wCen}
In the context of the AGB model \citep{dercole2008}, the high helium stars are  {\it stars born from pure AGB ejecta}. The helium abundance in the envelopes of massive AGB results from the ``second dredge up" occurring between the H--rich envelope of the giant and the H-exhausted core before the model climbs up the AGB, and is independent from all the input parameters (e.g. mass loss rates, convection efficiency) that affect the abundances of elements subject to  p--captures during the HBB phase. Standard models  for massive AGBs predict typical values for the helium abundance of Y=0.34--0.36. These values can be slightly different for different assumptions made in the models, so any formal ``discrepancy" with observations should not be considered as a reason to discard the model\footnote{Much more important indeed is the fact that the AGB model is the only one implying an {\it upper} limit close to these values for the helium abundance in the 2G stars, while for other models there is no such upper limit , and there is no evidence that stars with much larger initial helium populate the colour magnitude diagrams of GC stars \citep[see, e.g. the discussion by][for the massive rotating models]{chantereau2016}.}. In Fig.\,\ref{fig:2} we have formally extended this result to the CNO abundances, showing that some stars of \wCen, at different [Fe/H], have the CNO abundances of pure massive AGB ejecta.\\
Second generation star formation in the AGB model occurs in the GC's inner regions, where a cooling flow collects the AGB ejecta \citep{dercole2008}. The formation of stars with extreme chemical abundances from pure AGB ejecta must occur in a phase when the cluster's central regions are devoid of any gas with pristine chemical composition, out of which first-generation stars formed, before any further infall of external pristine gas occurs. There is one `standard' GC for which this clearly happened: NGC\,2808, where  both  the presence of a high--helium main sequence and  the extension of the horizontal branch stars \citep{piotto2007, dc2008} signal a population born from pure AGB ejecta. \\
{\it In our putative NSC \wCen,  the formation of extreme stars  must have occurred several times, at different metallicities}.  These multiple episodes of formation of extreme 2G stars in the core of \wCen\ could be linked to the action of the central black hole, periodically sweeping out the gas from its surrounding regions.  This hypothesis allows a qualitative analysis of the phases of star formation in \wCen\ in the framework of the AGB model.
If the gas in the central regions of the NSC, from time to time, is fully swept out, it  is initially replenished only by the winds of the AGB population of the NSC, allowing formation of extreme second generation stars, either in a burst or in a quiet star formation event. Subsequently infall of gas with pristine chemical composition will result in the formation of 2G stars out of AGB ejecta diluted with pristine gas and thus with intermediate chemical composition.  {\it If the infall event is massive}, large scale star formation occurs, and the new stars formed  will have a metallicity increased by the pollution of the local medium caused by the previous core collapse supernovae. \\
In the context of the AGB model we are presenting, the formation of the extreme populations at increasing [Fe/H] must be a recurring event,  because, at each time, AGB ejecta of different metallicities merge in a single cooling flow and mix, forming new populations with scarce spread in iron. Recent analyses of the chromosome maps show that in the galactic GCs where there is a measurable metallicity spread of the first generation stars, the 2G stars display a much smaller spread \citep{legnardi2022, latour2025}, a result consistent with what we expect if the 2G forms in a compact well mixed region in the cluster core. Thus, in \wCen,  the AGB stars contributing to the extreme populations of each given metallicity must be in a limited range of metallicities, so that their mixed iron abundances do not exceed the range. This may be the reason why the {\it lowest metallicity bins} do not host stars with extreme anomalies, e.g. stars with maximum helium difference with respect to the initial abundance \citep{clontz2025} or with maximum CNO cycling, as shown by the lack of points from \cite{marino2012cno} with [Fe/H]$<$--1.9 on the ``yellow strip" in Fig.\,\ref{fig:2}. 
In each [Fe/H] bin, stars with much larger [Fe/H] are not yet old enough to  contain evolving AGB masses. 
This simple argument allows to estimate a time lapse  \tdelay$\sim$100\,Myr till the occurrence of the major event forming stars of [Fe/H] in the range between --1.5 and --1.3.

\section{Full dating GS\_3073 and GN--z11 }
\label{sec5}
\subsection{Dating the stars losing winds with high N/O}
\label{modelmatch}
\begin{figure}
\begin{minipage}{0.95\textwidth}
\resizebox{.5\hsize}{!}{\includegraphics{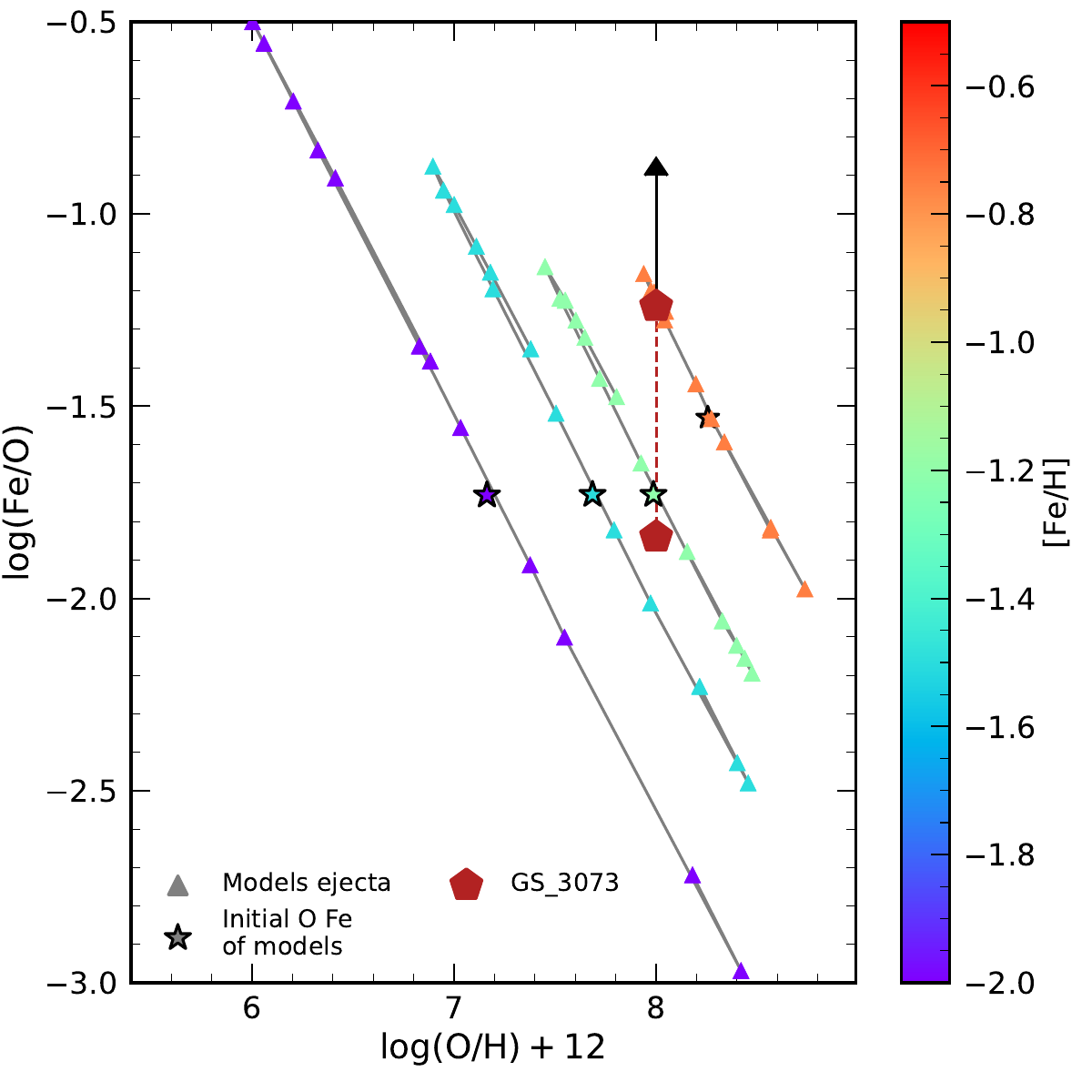}}	  
\end{minipage}
\caption{The range of lower limits for log(Fe/O)  for GS\_3073 are compared with the model ejecta values for the different sets of metallicities from Z=0.0003 (lower line) to Z=0.004 (upperline).}  
\label{fig:4}       
\end{figure}

\begin{figure}
\begin{minipage}{0.95\textwidth}
\resizebox{.5\hsize}{!}{\includegraphics{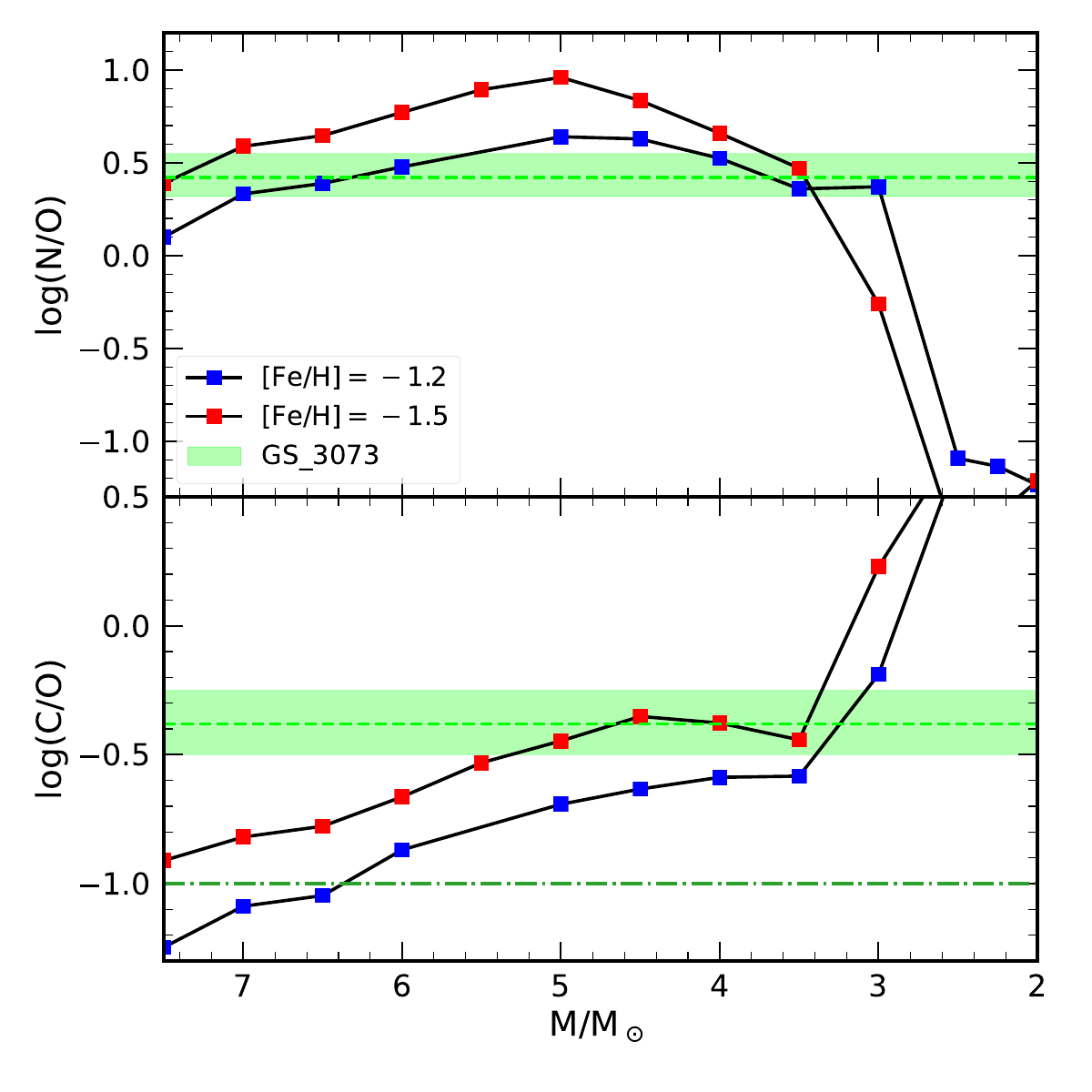}}	  
\end{minipage}
\caption{ Comparison of the abundance of ejecta as a function of the mass for [Fe/H]=--1.2 and --1.5. The fiducial abundance regions for GS\_3073 are the green strips, while the green dot-dashed line represents the lowest possible abundance for log(C/O).
 The evolving masses which best reproduce the fiducial abundances of the UV dense  region close to the SMBH are $\sim$4--3.5\Msun, for Z=0.001 and 4--3\Msun for Z=0.002.}  
\label{fig:5}       
\end{figure}

In Fig.\,\ref{fig:1} we see that the abscissa O/H is not a proxy of the metallicity of the models, but their O/H abundance in the ejecta: oxygen can be depleted, but also larger with respect to the initial abundance, depending on whether the specific  mass in evolution is in HBB (oxygen depleted) or is dominated by dredge up. If we are looking at pure ejecta, then {\it the specific metallicity of the AGB stars in play is not formally well fixed}.  
One option to constrain the Fe abundance relies on the comparison between the observed range of log[Fe/O] in  \cite{ji2024}  and the corresponding values from the models ejecta (Fig.\,\ref{fig:4}).
We can immediately exclude the very low metallicity (Z=0.0003, corresponding to [Fe/H]=--2 for our choice of $\alpha$\ elements overabundance) as the ejecta do not match any of the abundances, neither N/O nor  Fe/O nor O abundances. 
As we have discussed for the case of \wCen, we expect to be in a relatively late stage of the NSC evolution. where the supernovae Type\,Ia have begun exploding. 
When we refer to pure AGB ejecta, we refer to ejecta not diluted with pristine infalling matter  but contamination with the local iron ejecta of SN\,Ia is unavoidable. This implies that we can rule out that the AGB progenitors of the gas we see belong to the Z=0.004 models, corresponding to [Fe/H]=--0.75, although the model ejecta pass through the high fiducial lower limit of GS\_3073 (upper red pentagon). Such an Fe/O value implies that oxygen is approximately in solar proportions with respect to iron, that is, that the $\alpha$\ elements are not overabundant in the gas we are seeing.  But the system GS\_3073 at such an early age should have the $\alpha$\ elements overabundances of population II stars, unless it has been actually contaminated by SN\,Ia iron, as we say. Thus the AGBs responsible for the high N/O must have a smaller initial [Fe/H].\\
 We are left with the intermediate cases, Z=0.001 ([Fe/H]=--1.5) and 0.002 ([Fe/H]=--1.2). \\
In Figure\,\ref{fig:5} we compare the abundances of log(N/O) and log(C/O) of the ejecta for [Fe/H]=--1.5 and --1.2   as a function of the initial AGB mass, with log(N/O) and log(C/O) for GS\_3073 (green dashed regions). Considering at the same time both N/O and C/O fiducial points with their formal errors, the comparison suggests that we are looking at winds from the following initial mass ranges:

-- [Fe/H]=--1.5~~~~   M=4--3.5\Msun~  age: 170--230 Myr

-- [Fe/H]=--1.2~~~~   M=4--3.0\Msun~~~  age: 170--340 Myr

To these specific ages we must add the time \tdelay, elapsed from the formation of population 1 (which defines the birth of the seed BHs) till the infall episode giving origin to the progenitors of the AGB gas we are considering. For these times we have established lower limits in our schematic model for \wCen, and estimated at a  \tdelay$\gtrsim$100\,Myr\ the birth of the populations at  [Fe/H]=--1.5 and --1.3. Summing up the ages, the total age range is $>$270--330\,Myr for  [Fe/H]=--1.5 and $>$270--440\.Myr for   [Fe/H]=--1.3. 
In summary, we predict that the total age of GS\_3073 is 270--440\,Myr, and that this is the time interval during which the SMBH of this system accreted mass and reached the present day mass  of $\sim 1.6 \times 10^8$\Msun\ from its initial seed mass.

\subsection{Dating again GN-z11}
We now estimate the age of GN-z11 in the context of the theoretical framework presented in this paper.
In \cite{dantona2023} we have dated the system by assuming a single stellar population forming both the NSC and the BH
seed, simply because the N/O of GN-z11 requires dilution with pristine gas, and interpreted the observed abundances assuming the system was similar to a simple GC.  Nevertheless, the NSC in this system may be as complex as in \wCen, and the estimated ages of the evolving AGBs must be increased to account for the time spent from the birth of the first population of the NSC and the birth of the specific population which gave origin to the winds in question. Considering that intermediate populations in \wCen\ are present only up to [Fe/H]$\simeq$--1.6 \citep{clontz2025} a \tdelay$\sim$100\,Myr looks adequate, 
again in the hypothesis that the (larger) [Fe/H] of GN-z11 results from pollution by SN\,Ia.
The most typical age range of the first population AGBs is to 50--130\,Myr , so the age of the system increases to $\sim$150--230\,Myr. This would mean that the NSC formed $\sim$190--310\,Myr after the Big Bang, at z=17.7 -- 14.3. 

\begin{figure}
\begin{minipage}{0.98\textwidth}
\resizebox{.5\hsize}{!}{\includegraphics{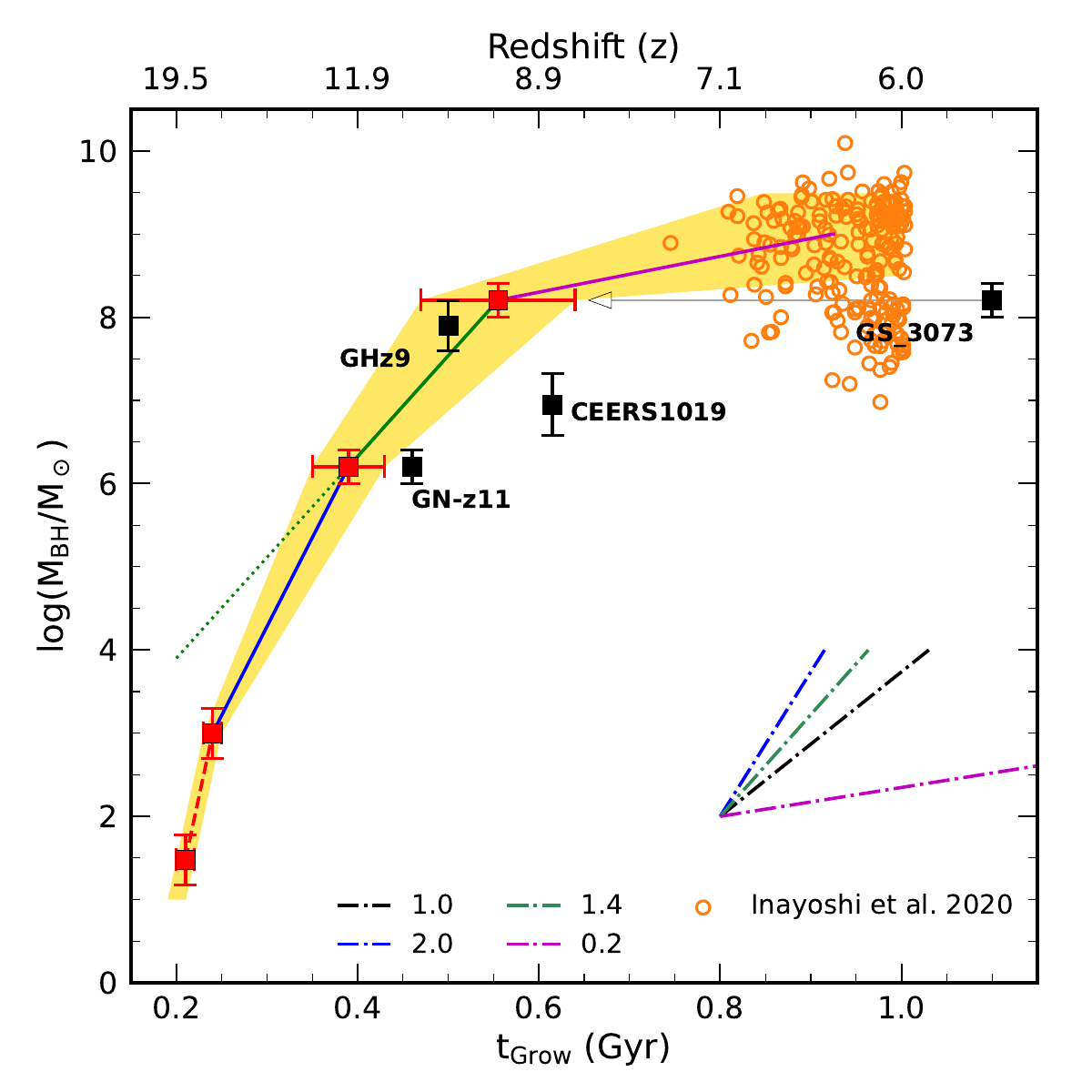}}	  
\end{minipage}
\caption{Schematic view of the growth of the seed BH through the GN-z11 and GS\_3073 masses up to the masses at z=6--7 \citep{inayoshi2020}. The black squares show the BH masses versus the age of the Universe corresponding to the redshifts of the four systems in Table\,\ref{table:1}.
We fix the zero age of the NSC formation at z=19.5 and age of the Universe of $\sim$200\,Myr.
The initial masses for the BH evolution are in the range allowed for stellar BH remnants (in this plot 30\Msun), and we assume dynamical merging to form a 1000\Msun\ seed during the first 30\,Myr after their formation (dynamical evolution is represented by the red dotted line).  We plot the systems GN--z11 and GS\_3073 at the ages found in this work (plus the 200\,Myr of the zero point of ages, red squares), based on the  N/O and C/O {\it vs.} O/H abundances of GS\_3073, and on the idea that they correspond to pure AGB ejecta, in a NSC system evolution resembling \wCen.  GN--z11 is obviously very close to its redshift age, while the horizontal arrow shows the shift to the `growing' age location of  GS\_3073. The accretion evolution is represented by the yellow band. The lines connecting  the 1000\Msun\ seed and the average locations of GN-z11, GS\_3073 and the quasars at z=6--7 give a qualitative idea of the mass growing rates. The slope for constant accretion rates equal to $\alpha$ times the Eddington mass accretion, with $\alpha$=0.2, 1, 1.4 and 2 are shown for a qualitative comparison with the slope of the segments connecting the objects. The dotted green line extrapolates back in time the line $\alpha \simeq$1.4 connecting the locations of GN-z11 and GS\_3073, and shows that an initial seed of $\sim 10^4$\Msun\ needs a uniform super-Eddington rate till the mass of GS\_3073. }   
\label{fig:6}       
\end{figure}

\subsection{Are AGB stars a plausible source of the N--rich gas? }
It is well known that the formation of second generation stars in ``standard" galactic GCs requires that the initial cluster mass is larger than the present day total mass. In the present case, we are dealing with structures which are not ending into standard GCs, at least because the deep gravitational well necessary to maintain into the clusters the interacting BHs also prevents to some extent the expulsion of supernovae ejecta, so that populations differing in metallicity are expected to be present. Further, we are dealing with high--z systems, where merging with other clusters or massive gas accretion from the surroundings is expected and, in fact, is needed to grow the mass of the central BH. Thus the mass budget does not pose a discriminant problem at this stage.\\
The mass of the N--rich ionized gas has been found to be $\sim1.2 \times 10^5$\Msun for CEERS-1019 \citep{marques2024} and $\sim 2 \times 10^5$\Msun\ for GN-z11 \citep{charbonnel2023}, although,  according to \cite{maiolino2024}, the latter value could be smaller. For GN-z11 the measured N/O may be the result of a mixture of AGB ejecta and infalling (standard N/O) gas. \cite{dantona2023} found that the abundances were compatible with 40\% ejecta diluted with 60\% infalling gas. On the contrary, CEERS-1019 lies close to the yellow strip of the pure AGB abundances in the plane log(N/O) vs. log(O/H) (see Fig.\ref{fig:1}) so its gas in our interpretation should be scarcely diluted.\\
Let us now ask what a mass of $\sim1.2 \times 10^5$\Msun\ implies if it were seen in GS\_3073. In this case, we have shown that it should come mainly from undiluted ejecta. We have also shown that the mass range 3-4\Msun\ is most probably contributing to this gas. Assuming a  \cite{kroupa2001} stellar initial mass function, with lower an upper limits equal to, respectively, 0.08 $M_{\rm \odot}$, 100 $M_{\rm \odot}$, stars in the range 3--4 $M_{\rm \odot}$\ provide an amount of AGB N-rich wind ejecta equal to about 3.3\% of the total stellar mass of the system. Consequently, an initial mass of of $3.6\times10^6$\Msun\ for the cluster providing these ejecta would be needed.

\section{Consequences on the mass accretion rates in GN--z11 and GS\_3073}
\label{tgrow}
We report the BH masses of the four objects listed in Table\,\ref{table:1} versus the age corresponding to their redshift (black squares) in Fig.\,\ref{fig:6}, where we also show the  location of the quasars at 6$<$z$<$7 from \cite{inayoshi2020}.
In order to understand the additional information coming from the dating of GN--z11 and GS\_3073 outlined in this work, 
we assume 200\,Myr (z=19.5) as formation epoch for the seed BHs, at and plot the growing ages as red squares. We start at 200\,Myr with initial stellar remnant seeds (located at 30\Msun\ in the figure) and assume that in 30\,Myr they form a seed of 10$^3$\Msun, starting point of the accretion phase. As we do not have any further age determination for GH-z9 and CEERS\,1019, we leave them at their redshift age. With the choice of starting the NSC formation at z=19.5, also GH-z9, at its redshift age, is in agreement with the growing time within the uncertainty band.  \\
We can compare the slope of the lines connecting the benchmarks with the standard growth timescales for accretion limited to the Eddington rate \citep{inayoshi2020}:
\begin{equation}
t_{\rm grow} \approx { {0.45 \epsilon} \over { (1-\epsilon) f_{\rm duty}}}  \ln \left({M_*}\over {M_{\rm seed}} \right) Gyr
\end{equation}
We take a duty cycle of accretion $ f_{\rm duty}$=1 and  a standard efficiency $\epsilon=0.1$. 
$M_{\rm seed}$\ is the seed (initial) mass, which we take as 10$^3$, $1.5 \times10^6$\ and $1.5 \times10^8$\Msun\ for the three phases of accretion delimited by the initial seed mass and by GN--z11 and GS\_3073 BH mass.
In Fig.\,\ref{fig:6} we see that growing the seed of 10$^3$\Msun\ up to the BH mass in GN-z11 requires a constant accretion at about twice the Eddington rate; slightly super--Eddington accretion is needed to reach the mass on GS\_3073, and from this value to the masses of the quasars at z=6--7 accretion can be largely sub-Eddington. Since the accretion rate is likely to be highly variable, the rates we find leave margin to form massive objects without invoking extreme accretion requirements.\\
This result compares well with the findings by \cite{schneider2023}:  in the cosmological context they find indeed that a system like GN--z11 should start from heavy seeds to grow up to the present mass without super--Eddington accretion phases, but it may start from light seeds, if it formed at z=20--24 and  was growing at super--Eddington rates.  \\
If BHs can sustain high rates of super--Eddington accretion, even intermittent, as proposed in several theoretical approaches \citep[e.g.][]{madau2014},
we can not exclude that even in GN-z11  the BH seeds were of standard stellar origin: 20--30\,\Msun\ remnants of the evolution of the massive stars formed together with the first stellar population. 
The stellar populations formed during the subsequent star formation events in the model also contribute to the growth of the central BH, by merging of newly formed stellar BHs or stars. Early merging with other star clusters, e.g. formed in the same dwarf galaxy \citep{garcia2025}, may further help to reach the requested mass in a timescale so short.
This, along with the longer epoch of star formation and BH accretion ($\sim 150-230$ Myr) that we suggest in the present work, may significantly reduce the global average accretion rate required for this system if the seed BHs are `light', remnant of standard stellar evolution \citep{maio2024gnz11, dantona2023}.

\subsection{The initial seed BH and its growth}
We focus again on the choice of the two first points (\Msun\ versus t$_{\rm grow}$ plotted in Fig.\ref{fig:6}. The ``standard" assumption made in this work is that the original BH seeds are 20--30\,\Msun, remnants of stellar evolution \citep{limongi2018}, that we are inside an NSC and that hierarchical repetitive mergers of these seeds lead to grow the BH mass, during the years preceding the end of the core collapse supernovae phase (a total of $\sim$30\,Myr) \citep[e.g.][]{antonini2019}, up to $\sim$1000\,\Msun. 
If clusters are so compact as those observed in the high redshift Universe, such as in the Sunrise Arc and the Cosmic Gems \citep{vanzella2023, adamo2024}, a much faster growth ---just a few Myr up to $\sim 10^3$\,\Msun--- of the {\it progenitors} of the seed BH  is possible, as found in recent simulations  \citep{laen2025} for clusters with stellar densities corresponding to very high surface densities (10$^5$\Msun/pc$^2$). In this model thousands of stars have mass $>$8\Msun, and several tens have mass $>$100\Msun. One of the merged stars has indeed M$>$1000\Msun and will collapse to a similar mass BH in a short time. Such massive remnants are also predicted by radiation hydrodynamic simulations of star-cluster formation, considering the interplay between metallicity, radiative feedback, and accretion dynamics in shaping the stellar populations in metal-poor environments \citep{chonomukai2025}, at least for metallicities [Z/H]=$10^{-2}$. The initial seed may then further grow by merging with the other stellar BHs remnants of the first cluster evolution, so that the mass at the second point in Fig.\ref{fig:6} may be even a factor 10 larger than our standard choice. Even with a very massive seed BH of 10$^4$\Msun, super--Eddington accretion would still needed to reach the mass of GN-z11, but the growth rate would be more similar to the rate between GN-z11 and GS\_3073.\\

\section{Conclusions}
\label{conclusions}
We have dated two N--loud AGNs in the young Universe under the requirement that the gas composition we see close to the BH comes from massive AGBs evolving in the Nuclear Star Cluster which originally gave birth to the seed BHs. We focused our attention on the observations by  \cite{ji2024}, who show that the N--loud region in the system GS\_3073 is confined to the central, dense and highly ionized regions, hosting an SMBH of $\sim 1.6 \times 10^8$\Msun, accreting at a sub--Eddington rates \citep{ubler2023}.
We show that the extreme log(N/O)=0.43 of this gas is consistent with the undiluted average abundances of AGB ejecta. A straight comparison with the N/O abundances in the galactic GC \wCen\ (likely the remnant NSC of a disrupted dwarf galaxy) shows that this abundance ratio matches those expected at a specific point of the evolution of the abundances in the `extreme' populations of the former nuclear star cluster \wCen. The analysis of the N/O, C/O and Fe/O abundances allow to constrain the age of the AGBs presently evolving in GS\_3073 to 170--340\,Myr. The presence, in \wCen, of extreme populations with distinct, increasing [Fe/H] and the hypothesis that they form in a cooling flow implicate a time delay of $\sim$100\,Myr from the formation of the NSC (and of the seed BHs) till the formation epoch of the progenitors of the AGBs presently losing the gas observed in GS\_3073. The final total age of the SMBH is then 270--440\,Myr.  A similar scheme applied to GN-z11 gives a total age of $\sim$150--230\,Myr. We do not have enough information for the other N--loud AGNs we listed, to constrain their age better than with the redshift age, but conclude that the age of the system GS\_3073 is much smaller than the redshift age. 
Including these two independent points in the diagram M$_{\rm BH}$\ versus age,  the BH growth with time in its first phases must be guided by phases of super-Eddington accretion. As shown in Fig.\,\ref{fig:6}, the rate decreases with aging, if the starting seed BH is $\sim 10^3$\Msun, made up by merging of stellar mass BHs during the first phases of the NSC evolution \citep{antonini2019}. Even assuming a more extreme initial seed of $\sim 10^4$\Msun, at the upper end of the VMS-SMS mass range predicted to emerge for the most compact NSCs \citep{fujii2024,laen2025}, it would be necessary to have a rate of accretion is $\sim$1.4 times the Eddington rate for the whole period until the age we determine for GS-3073.
After $\sim$300--500\,Myr, the rate becomes largely sub--Eddington. This result is in agreement with some theoretical predictions \citep[e.g.][]{madau2014}.\\
We conclude that the AGN -- AGB connection  may not only represent a direct explanation of the  N--loud AGNs, but, when linked with data detailed  enough on the system under exam, like in the case of GS\_3073, it may also add an independent estimate of the age of the system, a necessary step to look deeply at the puzzle of SMBH growth. 
The key point of the model is that accretion on the central BH, in its first growth, proceeds at intermittent phases of super--Eddington accretion, followed by quiescent phases in which accretion slowly resumes, first mostly from the NSC N--rich stellar winds, then by phases of massive infall, during which the AGB winds are first diluted (and the gas remains N--rich at a lower level) and then become an insignificant fraction of the accreting gas. Finding that about half of the young  N--rich galaxies have an AGN signature \citep{isobe2025} could be related to the relative timing of intermittent and accretion phases.
We have also theoretically defined a strip in the log(N/O) {\it vs.} log(O/H) plane which limits the pure AGB N--rich ejecta from our updated models \cite[see, e.g.][]{ventura2013}. In the proposed scheme of intermittent super Eddington accretion, we identify the objects on the strip as the AGN resuming accretion from ejecta, the objects with lower N/O (but larger than solar, at oxygen smaller then solar) as accreting from ejecta and infalling gas, while the AGN with standard N are objects as sources mostly accreting from infalling gas.
\\
Independently, we propose a scheme for the presence of multiple extreme ``second generation" stars in \wCen: we attribute their formation to the action of the central BH in sweeping out the gas from the core regions where these populations are then born from pure AGB cooling flow. 

\begin{acknowledgements} 
This work has been funded by the European Union -- NextGenerationEU RRF M4C2 1.1 (PRIN 2022 2022MMEB9W: ``Understanding the formation of globular clusters with their multiple stellar generations'', CUP C53D23001200006)." EV acknowledges support from NSF grant AST-2009193.\\
We thank Raffaella Schneider and Marco Limongi for useful insight on many facets of this work.
\end{acknowledgements}
\bibliographystyle{aa}
\bibliography{gs3073revised}
\end{document}